\documentclass[aps,pra,twocolumn,showpacs,superscriptaddress]{revtex4-1}
\usepackage{bbm}
\usepackage{graphicx}
\usepackage{amsmath}
\usepackage{amssymb,amsthm}
\usepackage{hyperref}
\usepackage[normalem]{ulem}
\usepackage{color}

\newcommand{\bra}[1]{\langle #1 |}
\newcommand{\ket}[1]{| #1 \rangle}
\newcommand{\braket}[2]{\langle #1|#2\rangle}
\newcommand{\mval}[1]{\langle #1 \rangle}
\newcommand{\eq}{\begin{eqnarray}} 
\newcommand{\en}{\end{eqnarray}}

\begin{document}

\title{Entanglement between two spatially separated ultracold interacting Fermi gases}

\author{P. Alexander Bouvrie}
\affiliation{Centro Brasileiro de Pesquisas F\'isicas, Rua Dr. Xavier Sigaud 150, Rio de Janeiro, RJ 22290-180, Brazil}

\author{Eloisa Cuestas}
\affiliation{Universidad Nacional de C\'ordoba, Facultad de Matem\'atica, Astronom\'ia, F\'isica y Computaci\'on, Av. Medina Allende s/n, Ciudad Universitaria, X5000HUA C\'ordoba, Argentina}
\affiliation{Instituto de F\'isica Enrique Gaviola (IFEG), Consejo de Investigaciones Cient\'ificas y T\'ecnicas de la Rep\'ublica Argentina (CONICET), C\'ordoba, Argentina}

\author{Itzhak Roditi}
\affiliation{Centro Brasileiro de Pesquisas F\'isicas, Rua Dr. Xavier Sigaud 150, Rio de Janeiro, RJ 22290-180, Brazil}
\affiliation{Institute for Theoretical Physics, ETH Zurich 8093, Switzerland}

\author{Ana P. Majtey}
\affiliation{Universidad Nacional de C\'ordoba, Facultad de Matem\'atica, Astronom\'ia, F\'isica y Computaci\'on, Av. Medina Allende s/n, Ciudad Universitaria, X5000HUA C\'ordoba, Argentina}
\affiliation{Instituto de F\'isica Enrique Gaviola (IFEG), Consejo de Investigaciones Cient\'ificas y T\'ecnicas de la Rep\'ublica Argentina (CONICET), C\'ordoba, Argentina}


\date{\today}

\begin{abstract}

Multiparticle entangled states, essential ingredients for modern quantum technologies, are routinely generated in experiments of atomic Bose-Einstein condensates (BECs). However, the entanglement in ultracold interacting Fermi gases has not been yet exploited. In this work, by using an ansatz of composite bosons, we show that many-particle entanglement between two fermionic ensembles localized in spatially separated modes can be generated by splitting an ultracold interacting Fermi gas in the (molecular) BEC regime. This entanglement relies on the fundamental fermion exchange symmetry of molecular constituents and might be used for implementing quantum applications in oncoming experiments. We show that the generated fermionic ensembles can be highly entangled and exhibit nonlocal quantum correlations. Entanglement-induced suppression of fluctuations in the single fermion spectral density of the ultracold fermionic gas is also observed.

\end{abstract}

\maketitle

\section{Introduction}

The progress towards the generation and manipulation of large ensembles of ultracold entangled atoms has been mainly focused on bosonic particles. Indeed,  most of the experiments aimed at generating multiparticle entangled states of matter, such as spin squeezing states \cite{EsteveOberthaler2008}, twin Fock states \cite{Gross2010,RiedelTreutlein2010,LuckeSchererEtal2011}, non-Gaussian states \cite{StrobelOberthaler2014,McConnellVuletic2015,Luo2017} or Dicke states \cite{LuckeKlempt2014}, deal with BECs. These states can exhibit full many-particle entanglement \cite{LuckeKlempt2014,Luo2017} including Einstein-Podolsky-Rosen (EPR) \cite{PeiseKlempt2015} and Bell \cite{TuraAcin2014,SchmiedSangouard2016} correlations. 

Although entanglement-enhanced precision in atomic interferometry has been achieved with the aforementioned states \cite{EsteveOberthaler2008,Gross2010,RiedelTreutlein2010,LuckeSchererEtal2011,StrobelOberthaler2014,McConnellVuletic2015,Luo2017,LuckeKlempt2014}, further quantum information applications require individual addressing of the subsystems. In addition, the indistinguishability of the atoms makes the standard notion of entanglement more subtle, since the very notion of entangled subsystems makes sense when each of the entangled parties can be individually addressed. Nevertheless, the generation of entanglement in identical particle systems is strongly related to the correlations due to the fundamental particle-exchange symmetry of the wavefunction. In particular, correlations appearing among inaccessible identical particles due entirely to symmetrization, can be extracted into an entangled state of independent modes in one-to one correspondence \cite{KilloranCramerPlenio2014,BouvrieValdesetall2016}. An example of this would be the splitting of a two indistinguishable-particle state into two individually addressable modes. For instance, the state $\left(\ket{ \uparrow \downarrow} \pm \ket{\downarrow \uparrow}\right)/\sqrt{2}$ yields the state $\left(\ket{\uparrow}_1 \ket{\downarrow}_2 \pm \ket{\downarrow}_1\ket{ \uparrow}_2 \right)/\sqrt{2}$, once each particle is fixed in one of the two modes. This entangled state (in the spin degrees of freedom) can be used for Bell measurements between two independent particle resources \cite{YurkeStoler1992}. Also, the generation of entanglement by splitting an ensemble of ultracold identical particles into two entangled twin-Fock states of an atomic BEC  was recently demonstrated \cite{LangeKlempt2018} and EPR steering has been observed \cite{Fadel2018,Kunkel2018}. The above procedure, which entangles individually addressable subsystems, will allow the exploitation of correlations due to indistinguishability as a resource in several quantum information tasks \cite{Cavalcanti2018}.

The Pauli exclusion principle makes the physics of ultracold interacting fermions and bosons to differ dramatically \cite{ChinGrimmEtal2010}. For instance, the crossover from BEC to BCS superfluidity \cite{ChenTanLevin2005,Giorgini2008}, a remarkable feature of strongly correlated fermion systems, is achieved with two-component ultracold interacting Fermi gases \cite{ZwierleinKetterle2004,BartensteinGrimm2004,BourdelKhaykovichEtal2004,PartridgeEtal2005,StewartJin2008}.
Since multiple occupation of the same single fermion state is forbidden, even the simplest state of identical fermions (a single Slater determinant) has correlations due to the fermion-exchange antisymmetry, which are extractable in the form of mode-entanglement \cite{BouvrieValdesetall2016}. In this regard, here we address the following questions. Is it experimentally possible to generate multi-particle entangled states by splitting an ultracold interacting Fermi gas? How strong is the generated entanglement? Does this entanglement  have observable consequences?  

Our goal is to demonstrate that large ensembles of fully entangled fermionic atoms can be generated with current technologies by splitting an ultracold interacting fermionic gas, and to give a complete set of theoretical tools to quantify the entanglement between the generated ensembles. We also demonstrate entanglement-induced suppression of fluctuations in the single particle-spectral density of strongly correlated fermionic gases and that the ensembles generated by splitting can be sufficiently entangled to exhibit nonlocal quantum correlations.

In the regime where the scattering length characterizing the interaction between different fermion species is positive ($a>0$), a finite fraction of fermion pairs condenses to the same molecular bound state $\ket{\psi_\textrm{gs}}$ forming a BEC of diatomic molecules \cite{GreinerRegalJin2003}. It has been shown that an ansatz of composite bosons (coboson) \cite{CombescotMatibetEtal2008} constitutes a good approximation for the ground state wavefunction at temperature $T=0$ on the BEC side of the crossover \cite{CombescotShiauChang2016,ShiauCombescotChang2016,BouvrieTichyRoditi2016}. From a Quantum Information point of view, the advantage of coboson theory over mean-field and Bogoliubov theories \cite{Giorgini2008,ChenTanLevin2005} lies in the fact that the many-body ground state can be written in terms of the single-fermion states of the system, and this allows the characterization of the many-particle quantum correlations present in the system and their observable consequences. 

By using the coboson ansatz we faithfully quantify the entanglement between two ensembles of fermionic atoms generated when splitting an ultracold interacting Fermi gas in the BEC side of the resonance. We predict that large ensembles -of the order of $10^5$ fully entangled fermionic atoms- can be generated in current experiments of ultracold interacting Fermi gases. Many-particle correlations in ultracold interacting Fermi gases increase when the system undergoes from the BEC to the unitarity regime. We also show that, close to the unitarity, the number of effective single-fermion states decreases to approximately the number of fermion pairs. This prevents fluctuations in the single-particle spectrum of the gas, experimentally observable in ultracold interacting Fermi gases \cite{StewartJin2008,Gaebler2010}. As for squeezed states of bosonic atoms, this suppression of fluctuations indicates strong entanglement among the fermionic atoms \cite{EsteveOberthaler2008}. When splitting the ultracold interacting Fermi gas, the resulting individually accessible fermionic ensembles can be highly entangled, and their single-particle spectral densities almost perfectly correlated.

Finally, we propose a Bell test of quantum nonlocality experimentally feasible only for a deterministic preparation of the state. We show that the CHSH inequality \cite{CHSH1969} can be violated by using quadrature phase amplitudes based on single-particle resolved measurements of the spectral density of the gas. For this ideal case, we demonstrate that the splitting of a strongly interacting Fermi gas generates two highly entangled fermionic ensembles with non local correlation and steering entanglement.

The article is organized as follows. We introduce the coboson ansatz and its application to two-component Fermi gases in section~\ref{SecCobAnsatz}. We thoroughly discuss the splitting process of an ultracold interacting Fermi gas and we analyze the particle correlation structure of the resulting state in section~\ref{SecEntGeneration}. In section \ref{SecEntQuant} we quantify the entanglement between the generated fermionic ensembles using the purity of the square reduced density matrix of one ensemble. In section~\ref{SecSpectDensFluc} we show how the entanglement is reflected as a suppression of fluctuations in the single-particle spectral densities of the fermionic ensembles. We present in detail the observables used for the Bell test in section~\ref{SecBellTest}, and discuss its validity and scope regarding the statistics and particle fluctuations of the prepared quantum state. Section~\ref{SecConclusions} is devoted to a summary and an outlook for future the work.

\section{Coboson ansatz for two-component Fermi gases}
\label{SecCobAnsatz}

The ansatz of composite bosons was simultaneously introduced in 2001 by A. J. Leggett \cite{Leggett2001} and by M. Combescot and C. Tanguy \cite{CombescotTanguy2001} for correlated pairs of bosonic and fermionic particles, respectively. In the case of fermions, it has been extensively applied to excitons \cite{CombescotShiau2015} which feature long-range Coulomb interactions. Its application to ultracold interacting Fermi gases was shown very recently: For an attractive short-range interaction between different fermion species ($A$ and $B$), it has been demonstrated that the ground state of two-component Fermi gases at zero temperature can be approximated by a Fock state of composite bosons, $\ket{N}$, whenever two-fermion bound states exist \cite{CombescotShiauChang2016}. Indeed, in the BEC regime, the universal dimer-dimer scattering length given by the coboson ansatz, $a_{dd}^\text{Cob}=0.64 a$ \cite{ShiauCombescotChang2016}, matches closely the well-established $a_{dd}\approx0.6a$ \cite{PetrovShlyapnikov2004}, and the molecular condensate fraction \cite{BouvrieTichyRoditi2016} matches remarkably well the fixed-node diffusion Monte Carlo \cite{Giorgini2005} and Bogoliubov results.

The pair-correlated state $\ket{N}$ is given by successive applications of identical {\it coboson operators} $\hat c^\dagger$ on the vacuum \cite{CombescotMatibetEtal2008},
\eq
\ket{N} =\frac{1}{\sqrt{N! \chi_{N}}} \left(\hat c^\dagger \right)^N \ket{0},
\en
where the operator $c^\dagger$ creates two fermions in a particular entangled state $\hat c^\dagger \ket{0} = \ket{\psi_{\textrm{gs}}}$. A well-known solution of $\ket{\psi_{\textrm{gs}}}$ for ultracold Fermi gases with large $N$ is the usual pair projection from a BCS state \cite{Nozieres1985,PongLaw2007}. Nevertheless, the solution given by the coboson ansatz is valid for any particle number $N$. Therefore it is possible to start from scratch by defining the operator $\hat c^\dagger$ such that $\ket{\psi_\textrm{gs}}$ is the ground state of a single-trapped-molecule \cite{BouvrieTichyRoditi2016}. Moreover, since any two particle system in a pure state admits always a state representation in terms of the single-particle states, the coboson state $\ket{N}$ can be written in terms of the single-fermion states of the system. Such state representation has two important advantages over the ground states used so far: It is rewarding when it comes to describe the particle correlations of the system and it allows to calculate observables based on the exact single-particle spectral density of the gas which is experimentally accessible \cite{StewartJin2008,Gaebler2010}.

Following \cite{BouvrieTichyRoditi2016}, we compute the molecular ground state $\braket{\vec r_\alpha, \vec r_\beta}{\psi_{\textrm{gs}}} = \psi_\textrm{gs} (\vec r_\alpha,\vec r_\beta)$ by solving the Schr\"odinger equation of a simple model of an harmonically trapped Feshbach molecule \cite{ChinGrimmEtal2010}. In particular, this is performed using a strong binding approximation between fermion A and B, and the model is applied to the ${}^6$Li broad resonance. The molecular state has six degrees of freedom associated with the coordinates of each particle $(r_\alpha,\theta_\alpha,\varphi_\alpha,r_\beta,\theta_\beta,\varphi_\beta)$. The interaction between fermion A and B introduces an extra symmetry (with respect to the non-interacting system) which increases the degeneracy of the system. The ground state is thus given by a particular superposition of all possible single-fermion states determined by the interaction strength and size of the trap.

By using the discretization technique of Refs. \cite{WangLawChu2005,KocikOkopiska2014,BouvrieTichyRoditi2016}, we carry out the Schmidt decomposition of the ground state,
\eq
\label{SchmidtRep}
\ket{\psi_{\textrm{gs}}} = \sum_n \sum_l \sum_{m=-l}^l \sqrt{\lambda_{nl}} \ket{\phi_{nlm}^{(\alpha)}(\vec r_\alpha)}\ket{\phi_{nlm}^{(\beta)}(\vec r_\beta)},
\en
where $\lambda_{nl}$ are the Schmidt coefficients and 
\eq
\braket{\vec r_{\gamma}}{\phi_{nlm}^{(\gamma)}(\vec r_{\gamma})} = \frac{u_{nl}(r_{\gamma}) Y_{lm}(\theta_{\gamma},\varphi_{\gamma})} {r_{\gamma}},
\en
with $\gamma=\alpha,\beta$ the Schmidt modes. That is, the state $\ket{\psi_\textrm{gs}}$ is written in the basis of the single-particle states associated to fermion A, $\{\phi_{nlm}^{(\alpha)}(\vec r_\alpha)\}$, and B, $\{\phi_{nlm}^{(\beta)}(\vec r_\beta)\}$, respectively. $Y_{lm}(\theta,\varphi)$ are spherical harmonic functions, and the radial functions $u_{nl}(r)$ are numerically obtained for a discrete space $r=(x_1,x_2,..., x_{max})$. The states $nlm$ have degeneracy $g_l=(2l+1)$ and, therefore, the single particle energies $E_{nl}$, as well as the Schmidt coefficients $\lambda_{nl}$, do not depend explicitly on the quantum number $m$. This is because the system is invariant under rotations around the axis $\vec r_\alpha - \vec r_\beta$. We order the $nlm$ states in increasing single-particle energy, and use one single-index $j$ to list these states such that 
\eq
\underbrace{\lambda_1}_{\underset{g_0=1}{nl=00}} < \underbrace{\lambda_2 = \lambda_3 = \lambda_4}_{\underset{g_1=3}{nl=01}}  < \underbrace{\lambda_5}_{\underset{g_0=1}{nl=10}} < \underbrace{\lambda_6 = \cdots = \lambda_{10}}_{\underset{g_2=5}{nl=02}}<\cdots 
\en
and $\ket{a_j} = \ket{\phi_{nlm}^{(\alpha)}(\vec r_\alpha)}$ and $\ket{b_j} = \ket{\phi_{nlm}^{(\beta)}(\vec r_\beta)}$. Note that the quantum number $n$ does not constrain the values of $l$, also because of the rotational symmetry. By using a single-index $j$, the ground state of Eq.~\eqref{SchmidtRep} reads
\eq
\label{SchmidtRepj}
\ket{\psi_\textrm{gs}}=\sum_{j=1}^S \sqrt{\lambda_j}  \ket{a_j}\ket{ b_j},
\en
with $\sum_{j=1}^S \lambda_j=1$ and $S$ the Schmidt rank. The computed Schmidt distribution $\Lambda =(\lambda_1,\lambda_2,\ldots,\lambda_S)$ depends on the ratio between the scattering length $a$ and the characteristic length of the trap $L$, $\Lambda  (a/L)$, and has finite but large enough $S$ (approximately equal to $10^6$). 

Given the Schmidt representation of the state \eqref{SchmidtRepj}, the coboson creation operator is naturally defined as $\hat c^\dagger = \sum_{j=1}^S \sqrt{\lambda_j}  \hat a^\dagger_j \hat b^\dagger_j$,  \cite{Law2005}, where $\hat a^\dagger_j$ ($\hat b^\dagger_j$) creates a fermion $A$ ($B$) in the single-fermion state $\ket{a_j}$ ($\ket{b_j}$).
Because of the Pauli principle, $( \hat a_{j}^\dagger)^2 = (\hat b_{j}^\dagger)^2=0$, and 
\begin{equation}
\label{NCobState}
\ket{N} = \frac{1}{\sqrt{N!\chi_N}} \sum^S_{\substack{j_1,j_2, \cdots, j_N = 1 \\ \sigma (j_1,\ldots, j_N)}} \left(\prod_{k=1}^N \sqrt{\lambda_{j_k}} \hat a_{j_k}^\dagger \hat b_{j_k}^\dagger \right) \ket{0},
\end{equation}
where $\sigma(j_1,\ldots, j_N)$ indicates that the sum over all the indices appearing in the summand has the restriction that the indices $j_1, j_2,\ldots,j_N$ must take distinct values. The $N$-coboson {\it normalization factor} $\chi_N$ \cite{Law2005,TichyBouvrie2012a,TichyBouvrie2014} is the elementary symmetric polynomial $\chi_{N} = N! \sum_{1<j_1 < j_2 < \cdots < j_N<S} \lambda_{j_1} \lambda_{j_2} \cdots \lambda_{j_N}$ and $\ket{0}\equiv \bigotimes_{j=1}^S \ket{vac}_{a_j} \otimes \ket{vac}_{b_j}$ is the vacuum. Note that the Pauli exclusion principle is guaranteed due to the restriction $\sigma(j_1,\ldots, j_N)$, i.e., there are not two labels with the same value. The ensemble of fermionic atoms $\ket{N}$ is controlled by the universal interaction parameter $k_Fa$, where $k_F = (6\pi^2 n)^{1/3}$ is the Fermi wave number of a non-interacting gas with atom-pair density $n = N/V$. The volume of the system is $V=4\pi L^3 /3$, with $L=\sqrt{\hbar/m_\text{Li} \omega}$ being the characteristic length of the trap, $\omega$ the confining frequency and $m_\text{Li}$ the atomic ${}^6$Li mass. 

The computed ground-state wave function of the fermionic ensemble $\ket{N}$ is only valid for $(k_Fa)^{-1} \ge 0.5$, in principle, because of the strong binding approximation performed to obtain the ground state of the Feshbach molecule $\ket{\psi_\textrm{gs}}$ \cite{BouvrieTichyRoditi2016}.  However, the impact of the confining potential on the many-body behavior, which becomes apparent close to the unitary regime $(k_Fa)^{-1} < 0.5$, is a difficult task to understand in ultracold fermionic gases \cite{Giorgini2008}. In general, the problem of non uniform configurations is highly non trivial. Semiclassical approximations, such as the local-density approximation, have provided a reliable and relatively simple description to infer the many-body behavior in large-size traps \cite{DalfovoStringari1999}. Therefore, coboson states $\ket{N}$ based on a more precise description of the pair correlated states $\ket{\psi'_\textrm{gs}}$ might be required to address harmonically trapping fermionic gases close to the unitary regime.

We emphasize that this representation of the state plays a key role for the improvement in computation time of the many-particle state (and observables) of interest when comparing to the computation time required by Quantum Monte Carlo (QM) techniques. In this work we deal with ensembles of $10^3$ fermionic atoms, which in fact it would be infeasible to simulate with other techniques such as QM or variational techniques \cite{Giorgini2005,BrouzosSchmelcher2013}. The required numerical simulations for computing the coboson state is feasible within short computational times \cite{WangLawChu2005,CuestasSerra2013,KocikOkopiska2014,BouvrieTichyRoditi2016}. Also, the analytical expressions used and developed here are relatively simple (in the sense that they do not require much quantum field knowledge), leading to results which include correlations neglected by the mean field approach, and also showing great agreement with the results obtained through Bogoliubov theory and QM simulations \cite{BouvrieTichyRoditi2016}.

\section{Generation of two entangled fermionic ensembles} 
\label{SecEntGeneration}

Here we show that two highly entangled fermionic ensembles can be generated by splitting an ultracold interacting Fermi gas in the BEC side of the resonance where $(k_Fa)^{-1} \ge 0.5$. We discuss the experimental conditions in which the splitting process should be performed and discuss the many-particle correlations of the resulting state

Beam-splitter-like dynamics in ultracold interacting Fermi gases can be very complicated to address theoretically. However, it has been shown for a few interacting fermions that in the strong-attractive-interaction regime, ultracold fermionic atoms cotunnel between two separated traps as pairs \cite{SerwaneZurnEtal2011,ZurnJochim2013,Jochim2015}. On the other hand, for a large number of fermion pairs, fluctuations between fermion hyperfine states are negligible when splitting an ultracold Fermi gas in the BEC regime into two spatially separated traps, thus keeping the fermionic ensembles of each trap unpolarized \cite{KohstallRiedlEtal2011}. Fermion pairs can therefore be described by a single bifermion creation operator $\hat d_j^\dagger=\hat a_j^\dagger \hat b_j^\dagger$, which simplifies the dynamics of the fermionic ensemble \cite{TichyBouvrie2012b,BouvrieTichyMolmer2016}. 

We consider a splitting dynamic  governed by the evolution operator $\hat d_{j}^\dagger \rightarrow ( \sqrt{R} \hat d_{1,j}^\dagger + \sqrt{T} \hat d_{2,j}^\dagger )$, where $R$ ($T=1-R$) is the reflection (transmission) coefficient, and $\hat d_{q,j}^\dagger = \hat a_{q,j}^\dagger \hat b_{q,j}^\dagger$ creates a fermion pair in the $j$th two-fermion state of mode $q=1,2$, i.e., $\hat d_{q,j}^\dagger \ket{0}_q = \ket{d_j}_q = \ket{a_jb_j}_q$. This unitary operation describes the experimental situation where a trapped fermionic gas is split into two identical traps of the same volume as the initial one, $V$, keeping the magnetic field fixed in order to preserve the interaction strength, the total particle correlation of the system and the value of global observables such as the total condensate fraction. Then the $N$-coboson Fock state evolves as \cite{BouvrieTichyMolmer2016} 
\begin{equation}
\label{FinStateWithOrthogonals}
\hspace{-0.3cm}
\ket{N} \longrightarrow \ket{\Psi_{N}} = \sum_{M=0}^N \sqrt{R^MT^{N-M}\binom{N}{M} } \ket{\Phi_{M,N-M}}, 
\end{equation}
where the states 
\begin{multline}
\label{FiState}
\ket{\Phi_{M,N-M}}= ( N! \chi_{N} )^{-\frac{1}{2}} \sqrt{ \binom{N}{M} } ~ \times \\ 
\times~\sum^S_{\substack{j_1,\ldots,j_{N} = 1 \\ \sigma(j_1,\ldots, j_N) }} \left(\prod_{k=1}^M \sqrt{\lambda_{j_k}} \hat d_{1,j_k}^\dagger \right) \left(\prod_{l=M+1}^{N} \sqrt{\lambda_{j_l}} \hat d_{2,j_l}^\dagger \right) \ket{0}_1 \ket{0}_2
\end{multline}
are orthonormal, $\braket{\Phi_{N_1,N_2}}{\Phi_{N_1',N_2'}}=\delta_{N_1,N_1'}\delta_{N_2,N_2'}$. For $(k_Fa)^{-1}<1$ the splitting dynamics of Eqs.~\eqref{FinStateWithOrthogonals} and \eqref{FiState} is jeopardized by molecular dissociations \cite{KohstallRiedlEtal2011}.

In Eq.\eqref{FinStateWithOrthogonals} we observe that fermion pairs are distributed binomially on the two modes of a perfect beam-splitter ($T=R=1/2$), as for ideal bosons or distinguishable particles. However, the final state $\ket{\Psi_{N}}$ is a multiparticle entangled state, since $\ket{\Phi_{M,N-M}} \neq \ket{M}_1\ket{N-M}_2$. Pauli correlations are preserved in the splitting process, see the constraints on the $j$'s in Eqs.~\eqref{NCobState} and \eqref{FiState}, and no more than a single fermion  occupies the same single-fermion state, independently of the mode in which it is localized. Analogous to the EPR thought experiment \cite{EinsteinPodolskyRosen1935}, measurements on ensemble $1$ yield predictions on the measurement results of ensemble $2$. Specifically, Pauli correlations between $M$ and $N-M$ fermion pairs in the initial state $\ket{N}$ are mapped onto multiparticle entanglement between two individual modes. This multiparticle entanglement between individual modes becomes operationally accessible when the system is projected onto the state $\ket{\Phi_{M,N-M}}$ with fixed particle number \cite{KilloranCramerPlenio2014,BouvrieValdesetall2016}. Nevertheless, the particle number in  both modes is usually determined during detection \cite{LangeKlempt2018} and therefore, even thought the prepared initial state has nonzero particle fluctuation, the system can be well described by states with a defined particle number.  

Quantum correlations generated by splitting a system of composite particles have already been discussed in Ref.~\cite{BouvrieTichyMolmer2016} where one of the generated ensembles (for instance 2) is brought to interfere with a third one (3). The entanglement between ensembles 1 and 2 is evidenced in the pair counting statistics between ensembles 2 and 3 after the interference process. Such interference scenarios are difficult to experimentally implement with molecular wave matter. In contrast with Ref.~\cite{BouvrieTichyMolmer2016}, in the present work we determine the entanglement between ensembles 1 and 2 generated by an experimentally feasible splitting process (without implementing any interference processes), as well as its behavior with the characteristic interaction parameter $k_Fa$ of ultracold interacting Fermi gases.

\section{Quantification of entanglement} 
\label{SecEntQuant}

The coboson ansatz $\ket{N}$ describes the many-particle system in terms of the single-fermion states $\ket{a_j}$ and $\ket{b_j}$. This allows us to faithfully describe the quantum correlations present in the system. Here, in order to quantify the entanglement between the generated fermionic ensembles we use the purity $P_q$ of one ensemble $q=1$ or $q=2$ ($P_1=P_2$). The reduced density matrix of ensemble $1$, $\rho_1$, is the partial trace of the density matrix $\rho = \ket{\Phi_{N,N-M}} \bra{\Phi_{N,N-M}}$ with respect to ensemble $2$,
\begin{equation}
\rho_1 = \sum_{1\le j_{1} < j_2 \cdots < j_{N-M} \le S} {}_2\bra{j_{1},\ldots,j_{N-M}}\rho \ket{j_{1},\ldots,j_{N-M}}_2,
\end{equation}
where
\begin{equation}
\label{BifermionState}
\ket{j_{1},\ldots,j_{n}}_q \equiv \prod_{k=1}^{n} \hat d_{q,j_k}^\dagger \ket{0}_q.
\end{equation}
By counting fermion states and their multiplicities, we show in Appendix~\ref{SecAppA} that the purity, 
\begin{equation}
P_1 = \sum_{1\le j_{1} < j_2 \cdots < j_{M} \le S} {}_1\bra{j_{1},\ldots,j_{M}}\rho^2 \ket{j_{1},\ldots,j_{M}}_1,
\end{equation}
is the symmetric polynomial 
\begin{equation}
\label{PurityBEC}
P_1 = \frac{1}{\chi_N^2} \sum_{\substack{j_1,\ldots,j_{N},i_1,\ldots,i_{N}=1 \\ \sigma(j_1, \ldots, j_{N}) \\ \sigma(i_1 , \ldots , i_{N}) \\ \sigma( i_{1}, \ldots, i_M ,j_{M+1}, \ldots, j_{N}) \\ \sigma(j_{1},\ldots, j_{M}, i_{M+1},\ldots, i_N) }}^S \prod_{k=1}^N \lambda_{j_k} \lambda_{i_k}.
\end{equation}
The numerical evaluation of Eq.~\eqref{PurityBEC} becomes infeasible for large $N$ and $S$, however, it can be expanded as a linear combination of elementary symmetric polynomials, 
\begin{equation}
\label{purity}
P_1= \binom{N}{M}^{-1} +   \sum_{m=0}^{N-2} \alpha_{m} \chi_{m}\chi_{2N-m}/ \chi_N^2, 
\end{equation}
where $\alpha_{m} = \alpha_{m}(N,M)>0$. Both $\alpha_{m}$ and $\chi_N$ can be evaluated by recursion formulas, allowing the computation of $P_1$ up to $N=10^3$, with $S\approx 10^6$. Note that the first term in Eq.\eqref{purity} constitutes a minimum for the purity, leading to the maximum entanglement that can be generated. It corresponds to the one given by the splitting of a single Slater state of $N$ identical fermions \cite{BouvrieValdesetall2016}, i.e. $P_1\geq \binom{N}{M}^{-1}$.

\begin{figure}[t]
\includegraphics[scale=0.4]{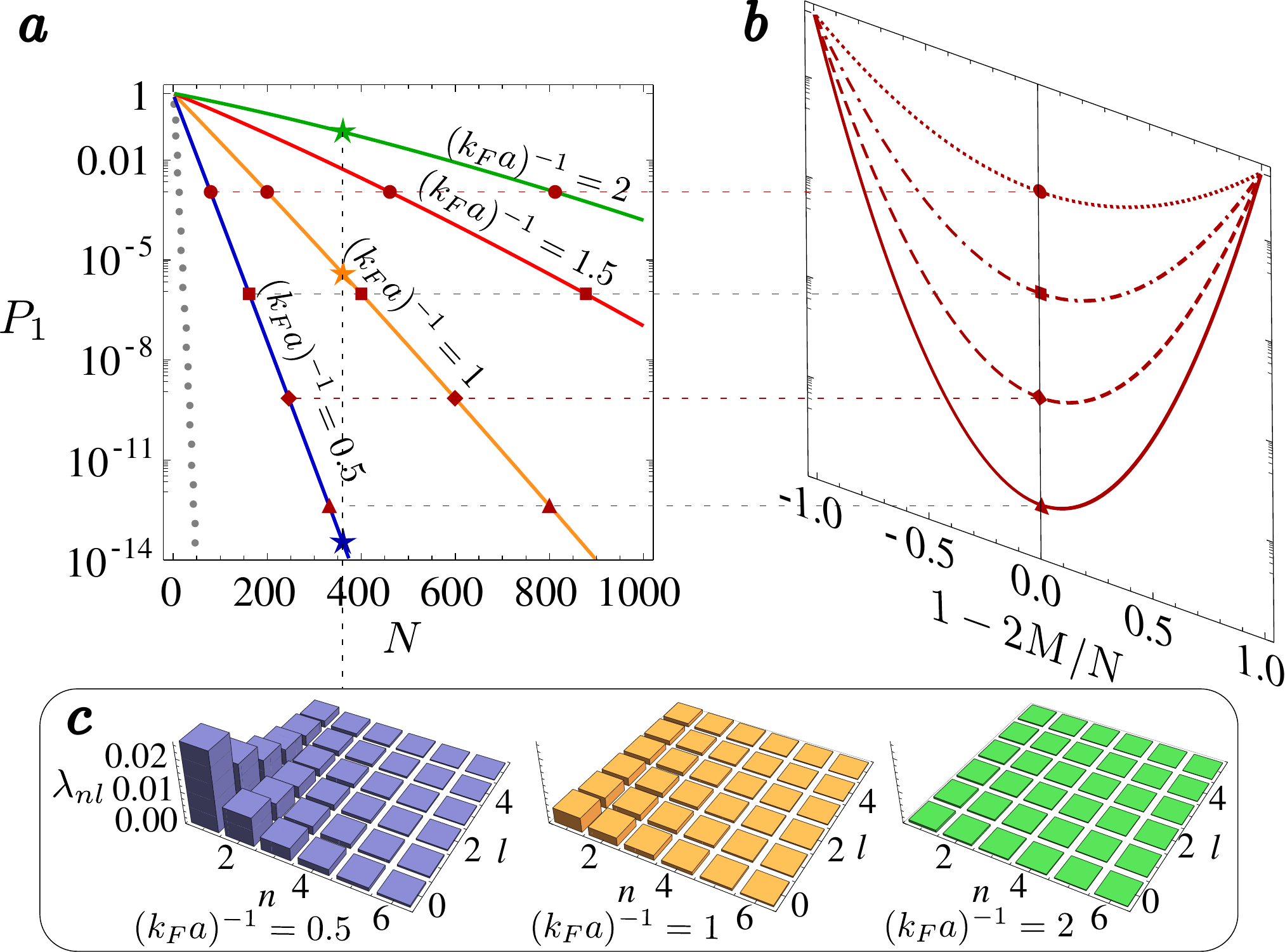}
\caption{$a)$ Purity of the fermionic ensemble $1$ (with $M=N/2$ fermion pairs) for different values of the interaction parameter $(k_Fa)^{-1}$. The gray dots represent the lower bound of the purity $P_1=\binom{N}{M}^{-1}$. For the red dots, squares, diamonds and triangles, respectively, the distribution of the purity with the population imbalance of the condensates ($1-2M/N$) is the same and it is shown in $b)$. $c)$ Schmidt coefficients $\lambda_{nl}$ for $N=360$ and $(k_Fa)^{-1}=0.5,1$, and $2$. We label single-particle states $(n,l)$ (with degeneracy $g_l=2l+1$) using a single index $j$. All depicted quantities are dimensionless.}
\label{PopuImbVsEnt}
\end{figure}

If molecular constituents are not perfectly bound, fermion exchange interactions become relevant yielding strongly correlated fermion ensembles \cite{Giorgini2008}. According to this, the entanglement between modes $(1|2)$ increases with $k_Fa$. We found that, for small $(k_Fa)^{-1}$ and large $N$, highly entangled mBECs are generated since the purity $P_1$ decreases many orders of magnitude (see Fig.~\ref{PopuImbVsEnt}$a$). 

The number of effective (non-negligible) Schmidt coefficients decreases as the value of $(k_Fa)^{-1}$ diminishes (see Fig.~\ref{PopuImbVsEnt}$c$). The competition for the occupation of the single-fermion states is the underlying reason for increasing correlations between fermion pairs. In the BCS limit ($(k_Fa)^{-1}\ll-1$) the momentum  distribution of the atoms vanishes for $k>k_F$ and the normalization ratio $\chi_{N+1}/\chi_N \xrightarrow[]{\text{BCS}} 0$ \cite{PongLaw2007}. Since the many-particle system has an infinite number of single-particle states, to fulfill the later limit for all $N$ the Schmidt distribution $\Lambda^\text{BCS}$ should have $S_\text{BCS}\le N$ effective Schmidt coefficient being the others $S-S_\text{BCS}$ fvinfinitesimally small. According to the above observations, we foresee that unitary ($(k_Fa)^{-1}= 0$) Fermi gases present a Schmidt distribution with just a few $S_\text{u} \gtrsim N$ effective coefficients. In this limit $\chi_{2N-m}\approx 0$ for $2N-m > S_\text{u}$ and therefore the entanglement generated approaches its maximum value with $P_1\approx \binom{N}{M}^{-1}$ (small gray dots in Fig.~\ref{PopuImbVsEnt}$a$). 

We observed also that the entanglement is equally distributed with the population imbalance of the condensates, independently of the interaction parameter $k_Fa$ and the total number of particles $N$. This is shown in Fig.~\ref{PopuImbVsEnt}$b)$, where we plot the purity as a function of $1-2M/N$. The behavior of the entanglement with the population imbalance is strongly related to the universality of ultracold interacting Fermi gases in the sense that both of them imply the constraint on the normalization ratio \cite{BouvrieTichyRoditi2016}
\begin{equation}
\left. \frac{\chi_{N+1}}{\chi_N} \right|_{k_Fa}  = \left.\frac{\chi_{N'+1}}{\chi_{N'}}\right|_{k_F'a'=k_Fa}, 
\end{equation}
where $k_F'/k_F = (N'/N)^{1/3}$.

\section{Fluctuations of single-particle spectral densities} 
\label{SecSpectDensFluc}

The coboson ansatz also has the advantage of providing an exact and closed expression for the single-particle spectral density of the fermionic gas. The single-particle density matrix associated with fermion A or B ($\rho_a$ or $\rho_b$) of the fermionic ensemble $\ket{N}$ has the same eigenstates ($\ket{a_j}$ or $\ket{b_j}$) as the two-fermion state $\ket{\psi_\textrm{gs}}$, which we have already obtained. Hence, the coboson ansatz allows us to find the exact single-particle spectral density
\begin{equation}
n_\text{spect} (nl) = g_l \left| \braket{N}{\phi_{nlm}^{(a/b)}(\vec r_{a/b})} \right|^2.
\end{equation}
In this section we show that the many-particle entanglement present in the system is reflected in the particle fluctuations of the single-particle spectral density.

The matrix elements of $\rho_a$ are $\bra{a_i} \rho_a\ket{a_j} = \mval{\hat a_i^\dagger \hat a_j}_N = N D_j[N] \delta_{i,j}$, where $\delta_{i,j}$ is the Kronecker delta, $D_j[N]=\lambda_j \chi_{N-1}^{\Lambda_j}/\chi_N$, and $\chi_{N-1}^{\Lambda_j}$ are the elementary symmetric polynomials of $\Lambda_j=(\lambda_1,\lambda_2,\ldots,\lambda_{j-1},\lambda_{j+1},\ldots,\lambda_{S})$. The eigenvalues of $\rho_a$ are therefore given by the diagonal elements of $\rho_a$, $\bra{a_i} \rho_a\ket{a_i}$, which are the populations of the single-fermion states of the gas. Such populations multiplied by the degeneracy $g_l$ give the exact single-particle spectral density $n_\text{spect} (nl)$. The energy of the single-fermion states is bound from below by the single-particle energy $E_1$ associated with $\lambda_1$ and from above by the continuum $E_\infty=0$. It is important to mention that the single-particle spectral density can be experimentally measured with an energy resolution of $\Delta E=h\times$2.1 kHz \cite{StewartJin2008,Gaebler2010}, coarser than the one required to measure with single-particle resolution ($\Delta E<E_2-E_1$). We also note that the pair density fulfills $\bra{d_j} \rho_d\ket{d_j} = \bra{a_j} \rho_a\ket{a_j}$, while $\bra{d_i} \rho_d\ket{d_j} \neq 0$ with $i\neq j$ due to fermion exchange correlations. 

\begin{figure}[t]
\includegraphics[scale=0.5]{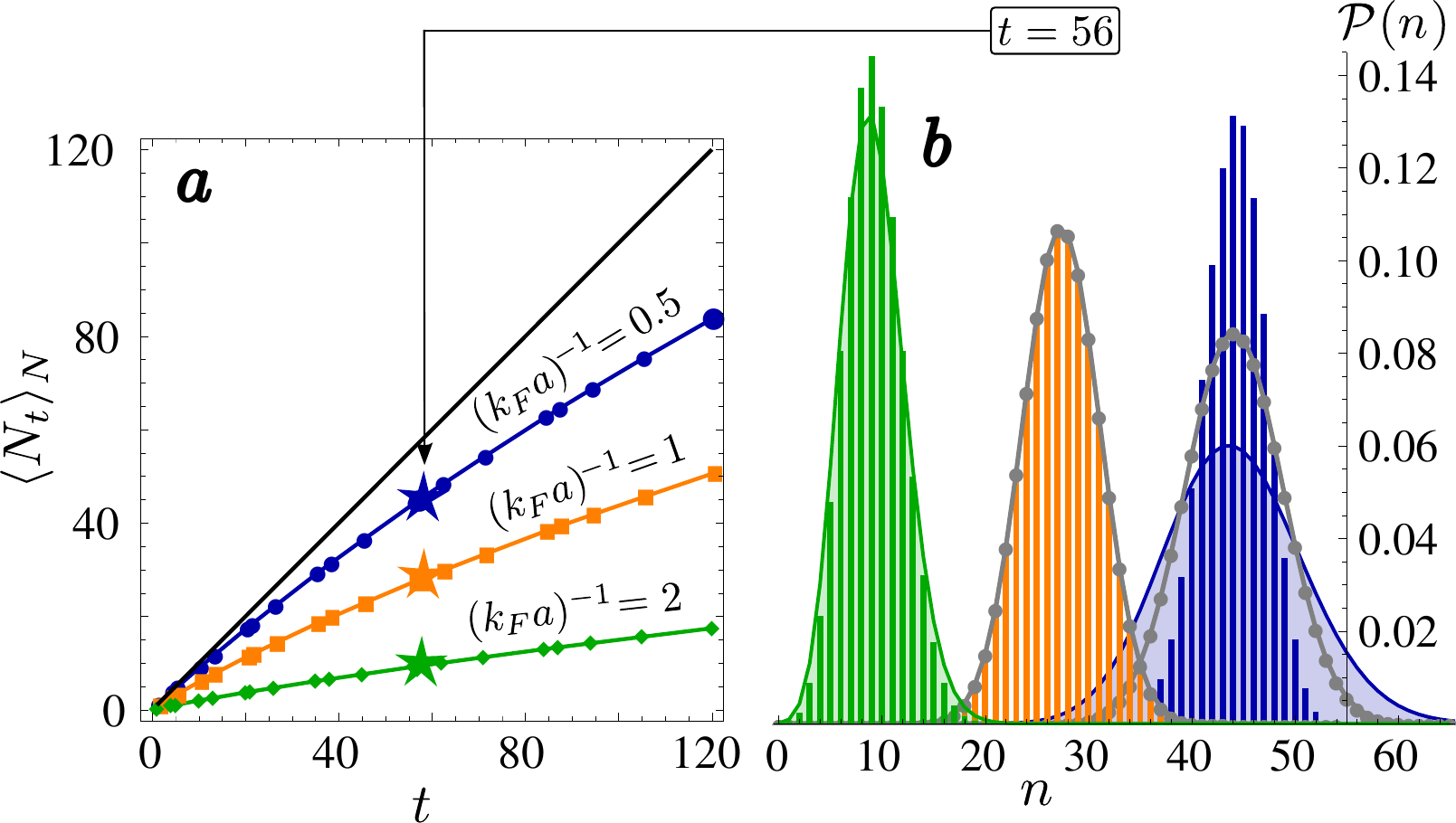}
\caption{$a)$ Mean population of the $t$-lowest energetic states of the single-particle spectrum ($\mval{N_t}_N = N\sum_{j=1}^t D_j$) of an interacting Fermi gas with $N=10^3$ fermion pairs, and interaction parameter $(k_Fa)^{-1}=2,1$, and $0.5$ (green, orange, and blue, respectively). Suppression of particle fluctuations in this spectral region $\tilde \Lambda_t$ is shown in $b)$ where the probability $\mathcal{P}(n)$ is plotted for $t=56$. Dashed areas are Poissonian distributions and connected gray dots are binomial distributions. All depicted quantities are dimensionless.}
\label{Figure2}
\end{figure}

Particle correlations can be read from the occupation probabilities of the single-fermion states, which give information on how the entanglement is distributed among the particles of the system. For instance, when $(k_Fa)^{-1} = 0.5$ (blue line), almost $t$ fermion pairs populate the $t$ lowest energetic states, $\tilde  \Lambda_t = (\lambda_1,\lambda_2, \ldots,\lambda_t)$, see Fig.~\ref{Figure2}$a$. Since no more than $t$ fermion pairs can populate this spectral region due to Pauli blocking, particle fluctuations are strongly suppressed. This is shown in Fig.~\ref{Figure2}$b$, where we plot the probability  $\mathcal{P}(n)=\sum_{1\le j_1<j_2< \cdots < j_n\le t} \mval{\prod_{k=1}^n \hat d_{j_k}^\dagger \hat d_{j_k}}_N = \binom{N}{n} \chi_{n}^{\tilde \Lambda_t} \chi_{N-n}^{\bar\Lambda_{S-t}}/\chi_N$, with $\bar \Lambda_{S-t} = (\lambda_{t+1},\lambda_{t+2}, \ldots,\lambda_S)$, of finding $n$ fermion pairs in $\tilde  \Lambda_t $. $\chi_{n>t}^{\tilde  \Lambda_t } = 0$ prevents populations larger than $t$. For $(k_Fa)^{-1}= 2$ (green) the fermionic ensemble behaves as a perfect BEC of uncorrelated bosonic molecules which yields Poissonian  distributions of $\mathcal{P}(n)$. Tuning to $(k_Fa)^{-1}= 1$ (orange) and then to $0.5$ (blue), the probability distribution $\mathcal{P}(n)$ changes from binomial to subbinomial. The latter resembles the typical particle distributions of spin squeezing states of entangled atomic BECs \cite{EsteveOberthaler2008}. Indeed, suppression of particle fluctuations in this regime is a direct consequence of the Pauli exclusion principle among identical fermions, a correlation of purely quantum nature. This demonstrates, therefore, entanglement-induced suppressions of particle fluctuations in ultracold interacting Fermi gases.

\begin{figure}[t]
\includegraphics[scale=0.8]{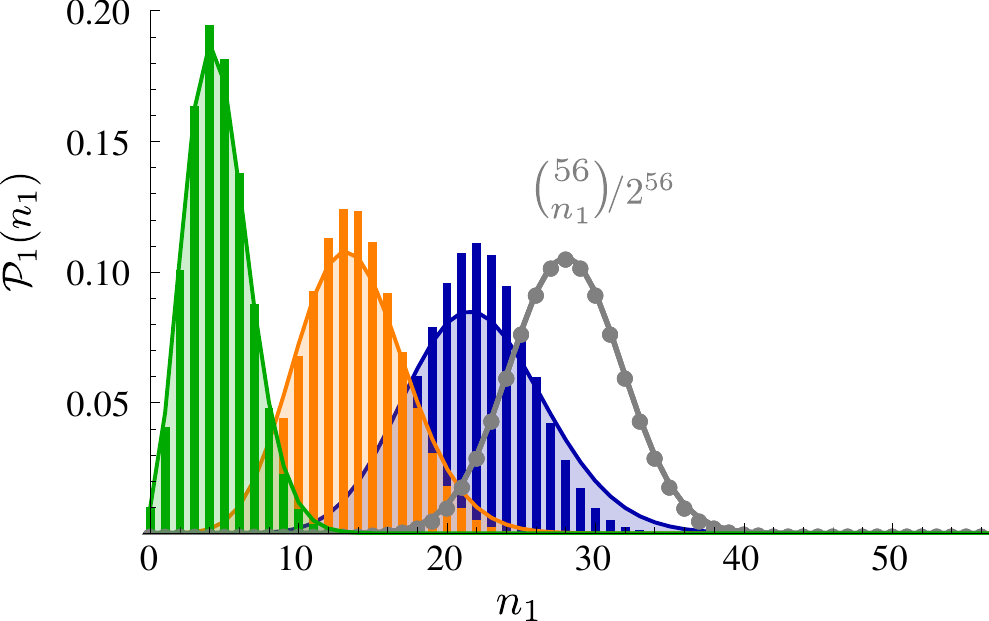}
\caption{Particle fluctuations in the states $\tilde \Lambda_t$ of the fermionic ensemble $1$ ($\mathcal{P}_1(n_1)$) after splitting the system into two balanced ensembles of $M=N/2$ fermion pairs. When decreasing $(k_Fa)^{-1}$, $\mathcal{P}_1(n_1)$ approaches the binomial distribution $\binom{t}{n_1}/2^{t}$ of two maximally entangled fermionic ensembles with perfectly correlated fluctuations. All depicted quantities are dimensionless.
}
\label{Figure3}
\end{figure}

Splitting the interacting fermion ensemble, the probability of detecting $n_1$ and $n_2$ fermion pairs in modes $1$ and $2$, respectively, on $\tilde \Lambda_t$ is given by $\mathcal{P}_{1,2}(n_1,n_2)= \binom{M}{n_1}\binom{N-M}{n_2} \chi_{n}^{\tilde \Lambda_t} \chi_{N-n}^{\bar\Lambda_{S-t}}/\chi_N$. If $(k_Fa)^{-1}$ decreases, the entanglement between these $n_1$ and $n_2$ particles moves towards its maximum value. This is reflected by the sub-Poissonian probability distribution $\mathcal{P}_1(n_1)=\sum_{n_2=0}^t \mathcal{P}_{1,2}(n_1,n_2)$ of finding $n_1$ fermion pairs in the spectral region $\tilde \Lambda_{t}$ of mode $1$ (Fig.~\ref{Figure3}). For $(k_Fa)^{-1} = 0.5$ and $t = 56$, $\mathcal{P}_1(n_1)$ approaches the binomial distribution $\binom{t}{n_1}/2^t$, and particle fluctuations in each individual mode are highly correlated in this spectrum range.  

\section{Nonlocal quantum correlations.} 
\label{SecBellTest}

Entanglement constitutes a fundamental resource for modern quantum technologies. However, most quantum applications require some kind of specific highly entangled states. For instance, quantum steering is used as a resource for secure quantum teleportation \cite{QiongyiReid2015}. According to Wiseman et al. \cite{WisemanDoherty2007}, there is a hierarchy of quantum correlations: The steerable states are a subset of the entangled ones and a superset of states exhibiting Bell non-locality. Here, we propose a Bell test of quantum nonlocality, which can be achieved experimentally with a deterministic preparation of the state. As we will explain later, it is also feasible to perform this test when the states present only particle number fluctuations. Nevertheless, it cannot be used to detect nonlocal quantum correlations in experiments with large ensembles of interacting fermions because the preparation of the quantum state is highly probabilistic. On the other hand, the proposed Bell test seems feasible for a few interacting fermions since the preparation of the state is now deterministic \cite{SerwaneZurnEtal2011}. From few to many fermion pairs, our results demonstrate that the entanglement between the generated ensembles can be large enough to present non-local quantum correlations and therefore steering entanglement, both essential resources for the implementation of quantum applications.

Bell-like quantum correlations can be recognized in the state $\ket{\Phi_{M,N-M}}$, Eq.~\eqref{FiState}, when it is written as a superposition of states having a fermion pair in the $j$th states of mode $1$  or $2$, respectively. The resulting equation is:
\eq
\label{FiStateBell}
\ket{\Phi_{M,N-M}} &=& \sqrt{M D_j} ~ \hat d^\dagger_{1,j} ~\ket{\Phi^{[\lambda_j]}_{M-1,N-M}} + \nonumber \\
&+& \sqrt{(N-M) D_j} ~ \hat d^\dagger_{2,j} ~ \ket{\Phi^{[\lambda_j]}_{M,N-M-1}} + \nonumber \\
&+& \sqrt{1-N D_j} ~ \ket{\Phi^{[\lambda_j]}_{M,N-M}}.
\en
where $\ket{\Phi^{[\lambda_j]}_{N_1,N_2}}$ is given by the state $\ket{\Phi_{N_1,N_2}}$ being removed from the terms with populated single-fermion state $j$. The state \eqref{FiState} becomes a maximally entangled Bell-like state when the occupation probability of the state $j$ fulfills $ND_j\to 1$. 

Inspired by Eq.~\eqref{FiStateBell}, we consider the quadrature phase amplitudes $\mathcal{Q}=Z_1$, $\mathcal{R}=X_1$, $\mathcal{S}=(X_2-Z_2)/\sqrt{2}$, and $\mathcal{T}=(X_2+Z_2)/\sqrt{2}$, based on projections of the single-particle spectral density with the single-fermion state $j$  of mode $q$ occupied 
\begin{equation}
\ket{o_{j,N_q}}_{q} =  \sum_{\substack{1\le j_1<j_2< \cdots< j_{N_q-1} \le S }} \hat d_{q,j}^\dagger \ket{j_1, \ldots, j_{N_q-1}}_q,
\end{equation}
or empty, 
\begin{equation}
\ket{e_{j,N_q}}_{q} = \sum_{\substack{1\le j_1<j_2< \cdots< j_{N_q+1} \le S }} \hat d_{q,j} \ket{j_1, \ldots, j_{N_q+1}}_q,
\end{equation}
namely, $Z_q=\sum_{N_q}\left(\ket{e_{j,N_q}}_{q~q}\bra{e_{j,N_q}}-\ket{o_{j,N_q}}_{q~q}\bra{o_{j,N_q}}\right)$, together with measurements in the rotated spectrum, $X_q=\sum_{N_q} \left(\ket{e_{j,N_q}}_{q~q}\bra{o_{j,N_q}}+\ket{o_{j,N_q}}_{q~q}\bra{e_{j,N_q}}\right)$. Classically, the possible results of the measurements in the ensemble $1$ ($2$) are $\mathcal{Q},~\mathcal{R}=\pm1$ ($\mathcal{S},~\mathcal{T}=\pm1$). Therefore, for local theories, the quantity $\mathcal{M}=\mathcal{QS}+\mathcal{RS}+\mathcal{RT}-\mathcal{QT}$ fulfills the CHSH inequality $\mathcal{M}\le 2$. Violation of the CHSH inequality would indicate the presence of nonlocal quantum correlations. The entanglement resulting from the splitting, rooted in the particle exchange symmetry, guarantees the existence of EPR correlations \cite{KilloranCramerPlenio2014,BouvrieValdesetall2016}, and the violation of the CHSH inequality $\mathcal{M}\leq 2$ \cite{CHSH1969} would demonstrate quantum nonlocality and, therefore, EPR steering.

\begin{figure}[t]
\includegraphics[scale=0.67]{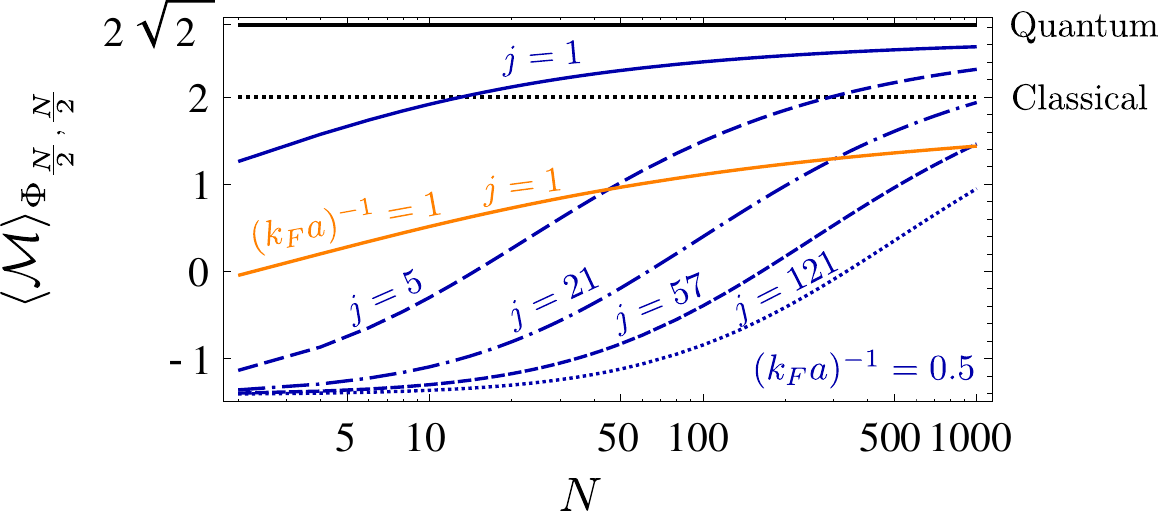}
\caption{Violation of the CHSH inequality ($\mathcal{M} \le 2$) for local theories. The represented states $j=1,5,21,57,121$ are the first nondegenerate states $(nl)$ of Fig.~\ref{PopuImbVsEnt}$c$ with $(k_Fa)^{-1}=0.5$, $l=0$ and $n=1,2,3,4,5$. All depicted quantities are dimensionless.}
\label{FigBelltest}
\end{figure}

For a deterministic preparation of the quantum state \eqref{FiState}, the result of the above measurements are 
\eq
\label{QS}
\mval{\mathcal{QS}}_{\Phi_{M,N-M}} &=& \sqrt{2} \left( N D_j[N]-\frac{1}{2}  \right. \\ & & + \left. \sqrt{M} \sqrt{D_j[N] (1-N D_j[N] )}\right), \nonumber \\ 
\mval{\mathcal{RS}}_{\Phi_{M,N-M}} &=& \sqrt{2} \left( \sqrt{M (N-M)} D_j[N]  \right. \\ & & - \left. \sqrt{(N-M)}\sqrt{D_j[N] (1- ND_j[N] )}\right) , \nonumber \\
\mval{\mathcal{RT}}_{\Phi_{M,N-M}} &=& \sqrt{2} \left( \sqrt{M (N-M)} D_j[N]  \right. \\ & & + \left.\sqrt{(N-M)}\sqrt{D_j[N] (1- ND_j[N] )}\right) ,\nonumber \\ 
\label{QT}
\mval{\mathcal{QT}}_{\Phi_{M,N-M}} &=& \sqrt{2} \left( \frac{1}{2} - N D_j[N]  \right. \\ & & + \left. \sqrt{M} \sqrt{D_j[N] (1-N D_j[N] )}\right),  \nonumber
\en
and the mean value CHSH quantity reads
\begin{equation}
\label{MCHSH}
\mval{\mathcal{M}}_{\Phi_{M,N-M}} = \sqrt{2} \left( 2 D_j[N] \left( N + \sqrt{M(N-M)} \right) - 1 \right).
\end{equation}
In figure \ref{FigBelltest} we show the resulting $\mval{\mathcal{M}}_{\Phi_{N/2,N/2}} = \sqrt{2} \left( 3 N D_j[N] - 1 \right)$ for two balanced condensates ($M=N/2$) as a function of the total number of particles $N$. It is found that $\mval{\mathcal{M}}_{\Phi_{N/2,N/2}}$ reaches values above 2 for small $(k_Fa)^{-1}=0.5$ and large $N>10^3$, for the lowest energetic states ($j<21$). Fermions in this range of the spectrum are therefore strongly entangled with each other and with the rest of the system. 

However, the quantum state produced in experiments with large ensembles of ultracold Fermionic atoms is usually a mixed state with fluctuating particle number, i.e., the initial state reads $\rho_\text{spl}= \sum_{N_1,N_2} \zeta_{N_1,N_2} \rho_{N_1,N_2}$ with $\sum_{N_1,N_2} \zeta_{N_1,N_2} = 1$. Contrary to the deterministic case, for mixed states $\rho_\text{spl}$ the error of the measured observables \eqref{QS}--\eqref{QT} cannot be estimated from measurements of the spectral density ($\Delta D_j[N]$) and the number of particles ($\Delta N$ and $\Delta M$). In the following we will show that the uncertainty of the measured observables can be estimated when the initial state has fluctuations only of the particle number and, as assumed above, when the splitting process is ideal, i.e., completely non adiabatic and isolated from its environment. 

In an ideal splitting process, one can consider the final state
\eq
\label{rhon1n2}
\rho_{N_1,N_2} &=& \sum_r \xi_r \ket{\Phi^r_{N_1,N_2}} \bra{\Phi^r_{N_1,N_2}},
\en
where $r$ runs over the possible statistical mixtures of the system before the splitting, with different distributions $\Lambda^{(r)}$ possible. The mean value of the CHSH observable $\mathcal{M}$ is now given by $\mval{\mathcal{M}}_{\rho_\text{spl}} =  \text{Tr} [\hat{\mathcal{M}} \rho_\text{spl}]$. 
Since $\zeta_{N_1,N_2}\xi_r \ge 0$ $\forall N_1,~N_2$ and $r$, also the maximum value of $\mathcal{M}$ is 
\eq
\text{max}[ \mathcal{M} ] = \sum_{N_1,N_2,r} \zeta_{N_1,N_2} \xi_r \text{max} \left[ \mathcal{QS} + \mathcal{RS} + \mathcal{RT} - \mathcal{QT} \right] = 2, \nonumber 
\en
for local theories. The splitting operation preserves the quantum correlations among fermion pairs and the occupation probabilities of the single-fermion states $\mval{\hat a_{j}^\dagger \hat a_{j}}_{\rho_\text{in}} = \mval{\hat a_{1,j}^\dagger \hat a_{1,j}}_{\rho_\text{spl}} + \mval{\hat a_{2,j}^\dagger \hat a_{2,j}}_{\rho_\text{spl}} \equiv \bar N \bar D_j$, where $\mval{\hat a_{q,j}^\dagger \hat a_{q,j}}_{\rho_\text{spl}} \equiv \bar N_q \bar D_j$. The number of particles in each spatial mode $q=1,2$ must be determined in each experimental realization and the number of realizations should be large in order to obtain accurately the mean particle numbers $\bar N_1 + \bar N_2 = \bar N$. By measuring the above occupation probabilities, and the mean particle numbers $\bar N_1 \pm \Delta \bar N_1$ and $\bar N_2\pm \Delta \bar N_2$, the error $\pm \Delta \bar D_j$ can be inferred.

\begin{figure}[ht]
\includegraphics[scale=0.82]{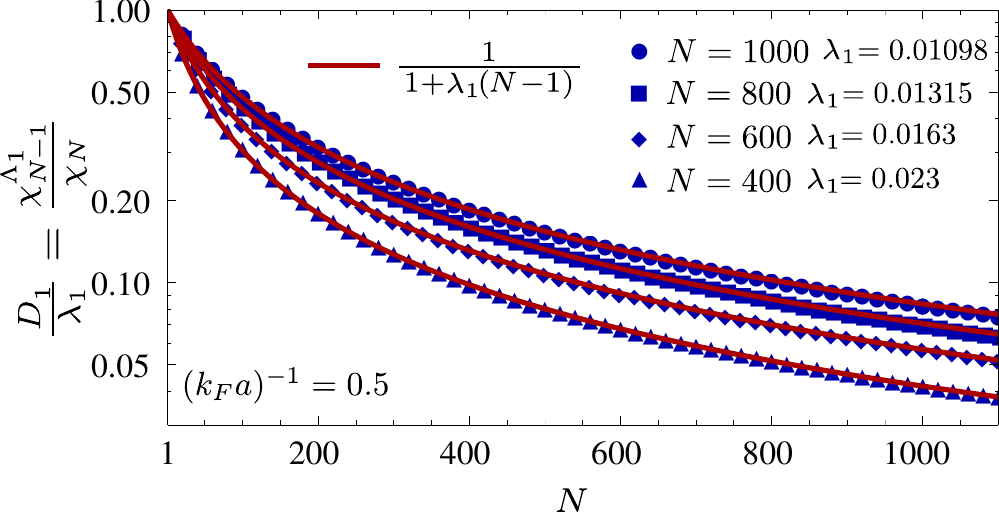}
\caption{Occupation probability of the lowest energetic state $N D_1[N]$, normalized to $N \lambda_1$, for a Fermi gas with $(k_Fa)^{-1}=0.5$ and a mean number of particles $\bar N$. Red lines are the function $1/(1+\lambda_j(N-1))$. All depicted quantities are dimensionless.}
\label{FigSupp}
\end{figure}

The error of the observables $\mval{\mathcal{QS}}$, $\mval{\mathcal{QT}}$, $\mval{\mathcal{RS}}$ and $\mval{\mathcal{RT}}$ can not be determined for the general state \eqref{rhon1n2} due to the mixedness of the single-fermion-states basis $\Lambda^{(r)}$.  Nevertheless, Fig. \ref{FigSupp} shows numerically that the spectral density factor $D_j[N]$ fits extremely well with the function $\lambda_j/(1+\lambda_j(N-1))$. Then, for any pure fluctuating particle number state, where $\Lambda^{(r)} \approx \Lambda$ for all $r$, the observables  $\mval{\mathcal{QS}}_{\rho_\text{spl}}$, $\mval{\mathcal{RS}}_{\rho_\text{spl}}$, $\mval{\mathcal{RT}}_{\rho_\text{spl}}$ and $\mval{\mathcal{QT}}_{\rho_\text{spl}}$ can be evaluated by replacing $N$, $M$, and $D_j$ by their mean values in Eqs.~\eqref{QS}-\eqref{QT},
\eq
\mval{\mathcal{QS}}_{\rho_\text{spl}} &\approx& \sqrt{2} \left( \bar N \bar D_j - \frac{1}{2} + \sqrt{\bar N_1} \sqrt{\bar D_j (1-\bar N \bar D_j )}\right),  \nonumber \\ 
\mval{\mathcal{RS}}_{\rho_\text{spl}} &\approx& \sqrt{2} \left( \sqrt{\bar N_1 \bar N_2} \bar D_j-\sqrt{\bar N_2}\sqrt{\bar D_j (1- \bar N \bar D_j)}\right) , \nonumber \\
\mval{\mathcal{RT}}_{\rho_\text{spl}} &\approx& \sqrt{2} \left( \sqrt{\bar N_1 \bar N_2} \bar D_j+\sqrt{\bar N_2} \sqrt{\bar D_j (1- \bar N \bar D_j)}\right) ,\nonumber \\
\mval{\mathcal{QT}}_{\rho_\text{spl}} &\approx& \sqrt{2} \left( \frac{1}{2} - \bar N \bar D_j  + \sqrt{\bar N_1} \sqrt{\bar D_j (1-\bar N \bar D_j )}\right),  \nonumber
\en
and their error can be estimated from $\Delta N_1$, $\Delta N_2$, and $\Delta D_j$. Considering that the spectral density of the statistical mixture of pure fluctuating particle number states is approximately $\bar N \bar D_j \approx N D_j $ for a given $(k_Fa)^{-1}<1$, the inequality $\mval{\mathcal{M}}_{\rho_\text{spl}} \le 2$ can be violated for nonzero particle invariance when $\bar N$ is large and $\bar N_1 \approx \bar N_2$. 

\section{Conclusions and outlook} 
\label{SecConclusions}

In the present work, by using an ansatz of composite bosons, we showed that many-particle entanglement between two fermionic ensembles localized in spatially separated modes can be generated by splitting an ultracold interacting Fermi gas. Interference with molecular wave matter was experimentally demonstrated in Ref. \cite{KohstallRiedlEtal2011} by splitting a mBEC of the order of $10^5$ fermion pairs. A double-well potential was generated by fast transforming a Gaussian optical dipole trap in order to keep the motional potential of the atoms, and thus almost reaching a perfect non adiabatic splitting. This splitting process was performed at large $(k_Fa)^{-1}\approx 3$, where our splitting dynamics apply. Then the external magnetic field, acting globally on both condensates, was adiabatically ramped down (in a time scale larger than $\omega^{-1}$ \cite{SerwaneZurnEtal2011}), increasing the scattering length. Since the interaction parameter $k_Fa$ can be tuned to arbitrary position, we obtained that two entangled fermionic ensembles with almost perfectly correlated spectral densities ($P_1\approx \binom{N}{M}^{-1}$) could be generated close to the unitary regime ($k_Fa<1$). The generated fermionic ensembles became strongly entangled in their single-fermion spectra and these EPR correlations can be large enough to exhibit quantum nonlocality. We showed violations of the CHSH inequality by using quadrature phase amplitudes based on single-particle state projective measurements, which demonstrated nonlocal quantum correlation and steering entanglement between particles of both ensembles. Violations involving states $j>1$ of both modes indicated that all fermions with single-particle energy lower than the energy $E_j$ will be even more entangled.

Although imperfect splitting processes can contribute to the mode mixing of both condensates destroying entanglement, close-to-ideal splitting is feasible in current experiments of ultracold bosonic \cite{Cavalcanti2018,Fadel2018,Kunkel2018,LangeKlempt2018} and fermionic \cite{KohstallRiedlEtal2011} gases. Entanglement-induced suppression of particle fluctuations can be detected experimentally if the energy resolution of the measured single-particle spectral density is increased to reach an error $\Delta E$ below $E_2-E_1$. Measurements of quadrature phase amplitudes, however, constitute an experimental challenge to be reached in ultracold Fermi gases. 
Also the state generated in the experiments concerning large ensembles of ultracold atoms can be highly probabilistic and then nonlocality could not be verified with the proposed Bell test. 

Interacting Fermi gases close to the unitary regime have been proven experimentally to be large ensembles of highly correlated fermions \cite{ChenTanLevin2005,Giorgini2008,StewartJin2008}.  These quantum correlations may derive in entanglement useful for most quantum information applications when the fermionic ensemble is split into two (Alice and Bob) or more individually accessible subsystems. The universal behavior of pair correlations in ultracold interacting Fermi gases might be useful to find entanglement witnesses for arbitrary mixed states based on macroscopic observables, since it allows one to relate the contact interaction between fermions at large momentum ($k>k_F$) to thermodynamic quantities experimentally accessible \cite{StewartEtal2010,KuhnleVale2010}. This, combined with alternative Bell inequalities involving only two-body correlations \cite{TuraAcin2014}, could allow one to implement tests of quantum nonlocality in real experiments with fermionic gases. Beyond the presented creation of spatial entanglement, Pauli correlations can be used to generate highly entangled fermionic ensembles in two spatially separated modes by using the interference between independent particle resources \cite{TichyBouvrie2012b}, or in many modes by separating the gas into single-molecules in an optical lattice. 

Finally, considering that interacting fermion systems of up to ten pairs were deterministically prepared in a quasi-one-dimensional dipole trap \cite{SerwaneZurnEtal2011,ZurnJochim2013}, where the number of available single fermion states is fully controlled, we expect that deterministic entanglement can be generated by splitting these interacting few-fermion system. Although in this work we focus on large ensembles of fermionic atoms, the coboson ansatz can also be applied to few-fermion systems in one dimension, for which the control of the quantum state advances impressively towards resolving quantum correlations \cite{BergschneiderPreiss2018}. The understanding of the entanglement in these one-dimensional few-interacting-fermion systems is particularly relevant, since the single-fermion states of the system are nondegenerated and can have almost unit probability, such that $S= N$, in the non interacting regime. Therefore, the maximally entangled state for each fermion species could be generated deterministically by performing splitting processes.

\begin{acknowledgements}

We thank Klaus M\"olmer, Malte Tichy, and Fernando de Melo for helpful discussions.  We also thank J.I. Robledo for his careful reading of the manuscript and the CSIRC of the University of Granada for providing the computing time on the Alhambra supercomputer. P.A.B. and I.R gratefully acknowledge support from the Conselho Nacional de Desenvolvimento Cient\'ifico e Tecnol\'ogico do Brasil and from the Spanish MINECO project FIS2014-59311-P (co-financed by FEDER). E.C and A.P.M. acknowledge the Argentinian agencies SeCyT-UNC and CONICET for their financial support. 

\end{acknowledgements}

\appendix

\section{Derivation of the purity $P_1$.} 
\label{SecAppA}

In order to quantify the amount of entanglement between two molecular BECs we calculate the purity $P_q$ of the reduced density matrix of the particles localized in one of the modes, for instance mode $q=1$. In the following appendix we derive analytically the purity $P_1$, i.e. Eq.\eqref{purity} in the main text.

Let $\rho_1$ be the reduced density matrix of particles in mode $1$ corresponding to the projected state $\ket{\Phi_{N,M}}$ (Eq. (3) in the main text) with a fixed number of particles in each mode, i.e.
\begin{widetext}
\eq
\label{ReducedDens}
\rho_1 = \text{Tr}_2 \left( \rho \right) = \sum_{1\le j_{M+1} < \cdots < j_{N}\le S} {}_2\bra{j_{M+1},\ldots,j_{N}}\rho \ket{j_{M+1},\ldots,j_{N}}_2,
\en
where $\rho = \ket{\Phi_{N,M}}\bra{\Phi_{N,M}}$, $\text{Tr}_2$ stands for the trace over all ($N-M$) particles in mode 2, and
\eq
\ket{j_{1},\ldots,j_{n}}_q = \prod_{k=1}^{n} \hat d_{q,j_k}^\dagger \ket{0}_q.
\en
Therefore, the reduced density matrix squared ($\rho_1^2$) reads
\eq
\rho_1^2 = \sum_{\substack{1\le j_{M+1} < \cdots < j_{N}\le S \\ 1\le i_{M+1} < \cdots < i_{N}\le S}}  {}_2\bra{j_{M+1},\ldots,j_{N}}\rho \ket{j_{M+1},\ldots,j_{N}}_{2~2}\bra{i_{M+1},\ldots,i_{N}}\rho \ket{i_{M+1},\ldots,i_{N}}_2.
\en
The purity of the reduced density matrix is given by the trace of $\rho_1^2$,  
\eq
P_1=\text{Tr}(\rho_1^2) = \sum_{1\le j_{1} < \cdots < j_{M}\le S} {}_1\bra{j_{1},\ldots,j_{M}}\rho_1^2 \ket{j_{1},\ldots,j_{M}}_1.
\en

Projections of the state $\ket{j_{M+1},\ldots,j_{N}}_2$ onto $\ket{\Phi_{N,M}}$ are straightforwardly obtained by counting the multiplicity of $\ket{j_{M+1},\ldots,j_{N}}_2$, that is,
\eq
\label{BifermionsProyection}
{}_2\braket{j_{M+1},\ldots,j_{N}}{\Phi_{N,M}} = (N! \chi_{N})^{-1/2} \sqrt{\binom{N}{M}} (N-M)! \left(\prod_{j=M+1}^{N} \sqrt{\lambda_{j_j}}\right) \sum_{\substack{j_1,\ldots,j_{M}=1 \\ \sigma(j_1, \ldots, j_{N})}}^S \left( \prod_{k=1}^M \sqrt{\lambda_{j_k}} \hat d_{1,j_k}^\dagger \right) \ket{0}_1.
\en
From Eq.~\eqref{BifermionsProyection} it follows that
\eq
\braket{\Phi_{N,M}}{j_{M+1},\ldots,j_{N}}_2 {}_2\braket{i_{M+1},\ldots,i_{N}}{\Phi_{N,M}} = \frac{(N-M)!}{\chi_N} \prod_{k=M+1}^N \sqrt{\lambda_{j_k}\lambda_{i_k}} \sum_{\substack{i_1,\ldots,i_{M}=1 \\ \sigma(i_1 ,\ldots, i_{N}) \\ \sigma( i_1 ,\ldots, i_{M}, j_{M+1}, \ldots, j_N) }}^S \prod_{l=1}^M \lambda_{i_l},
\en
and 
\eq
{}_1\bra{j_{1},\ldots,j_{M}}{}_2\braket{i_{M+1},\ldots,i_{N}}{\Phi_{N,M}} = (N! \chi_{N})^{-1/2} \sqrt{\binom{N}{M}} M! (N-M)! \left(\prod_{k=1}^{N} \sqrt{\lambda_{j_k}}\right) \left(\prod_{l=M+1}^{N} \sqrt{\lambda_{i_l}}\right),
\en
from which the following symmetric polynomial is obtained:
\eq\label{infsum}
P_1 &=& \frac{1}{\chi_N^2} \sum_{\substack{j_1,\ldots,j_{N},  i_1,\ldots,i_{N}=1 \\ \sigma(j_1, \ldots, j_{N}) \\ \sigma(i_1, \ldots, i_{N}) \\ \sigma(i_{1}, \ldots, i_M, j_{M+1}, \ldots, j_{N}) \\ \sigma(i_{M+1}, \ldots, i_N, j_{1}, \ldots, j_{M})}}^S \prod_{k=1}^N \lambda_{j_k} \lambda_{i_k}.
\en
\end{widetext}

For a large number of particles $N\gg 1$ and Schmidt coefficients $S>N$, the numerical evaluation of the sum in Eq.~\eqref{infsum} becomes infeasible. However, we can expand such equation as a linear combination of elementary symmetric polynomials $\chi_N$,  which can be evaluated for large $N$ and $S$ by using the recursion \cite{RamanathanKurzynski2011}
\eq
\label{ChiRecursion}
\chi_N=(N-1)!\sum_{m=1}^N \frac{(-1)^{m+1}}{(N-m)!} M(m) \chi_{N-m},
\en
where $M(m)=\sum_{j=1}^S \lambda_j^m$ are the power sums \cite{CombinatoryAnalysisMacmahon} of the Schmidt coefficient distribution ${\bf \Lambda }= (\lambda_1,\ldots,\lambda_S)$. Eq.~\eqref{ChiRecursion} allows us to evaluate the purity $P_1$, e.g., with $N\approx10^3$ and $S\approx 10^6$. Since Eq. \eqref{infsum} is a symmetric polynomial containing $2N$ coefficients with $2$ as maximum multiplicity, we can perform the partial sum of indices $j_{M+1}\dots j_N$ and $i_{M+1}\dots i_N$. By counting the multiplicity of the coefficients, Eq.~\eqref{infsum} can be written as

\begin{widetext}
\eq
P_1 &=& \frac{1}{\chi_N^2} \sum_{L=0}^{M} \frac{1}{L!}\left(\frac{M!}{(M-L)!}\right)^2 \sum_{\substack{j_1,\ldots,j_{2M-L}=1 \\ \sigma(j_1, \ldots , j_{2M-L})}}^S \left(\chi_{N-M}^{[j_1,\ldots,j_{2M-L}]}\right)^2 \prod_{k=1}^{L} \lambda_{j_k}^2 \prod_{l=L+1}^{2M-L} \lambda_{j_l}, 
\en
where $[j_1, ... , j_p]$ is the Schmidt coefficient distribution ${\bf \Lambda}=(\lambda_1, \lambda_2, ..., \lambda_S)$ without considering the coefficients $\lambda_{j_1}, ... , \lambda_{j_p}$. If we take into account the relation \cite{BouvrieTichyMolmer2016}
\eq
\label{ChiChiExpansion}
\chi_{N_1}\chi_{N_2} = \sum_{L=0}^{N_1}\frac{N_1! N_2!}{L! (N_1-L)! (N_2-L)!}
\sum_{\substack{j_1,\ldots,j_{N_1+N_2-L}=1 \\ \sigma( j_1 , \ldots , j_{N_1+N_2-L})}}^S \left(\prod_{k=1}^{L} \lambda_{j_k}^2 \right) \left(\prod_{l=L+1}^{N_1+N_2-L} \lambda_{j_l} \right),
\en
with $N_2 = N_1 = N-M$, in order to expand $\left(\chi_{N-M}^{[j_1,\ldots,j_{2M-L}]}\right)^2$, we obtain
\eq
P_1 &=& \frac{1}{\chi_N^2} \sum_{L_1=0}^{M} \sum_{L_2=0}^{N-M} \frac{1}{L_1!}\left(\frac{M!}{(M-L_1)!}\right)^2 \frac{1}{L_2!}\left(\frac{(N-M)!}{(N-M-L_2)!}\right)^2  \nonumber \\
& &\times \sum_{\substack{j_1,\ldots,j_{2N-L_1-L_2}=1 \\ \sigma(j_1, \ldots, j_{2N-L_1-L_2})}}^S \prod_{k=1}^{L_1+L_2} \lambda_{j_k}^2 \prod_{l=L_1+L_2+1}^{2N-L_1-L_2} \lambda_{j_l} \\
&=& \frac{1}{\chi_N^2} \sum_{L_T=0}^{N} \sum_{L_1=\text{Max}[0,L_T-N+M]}^{\text{Min}[L_T,M]} \frac{1}{L_1!}\left(\frac{M!}{(M-L_1)!}\right)^2 \frac{1}{(L_T-L_1)!}\left(\frac{(N-M)!}{(N-M-L_T+L_1)!}\right)^2  \nonumber \\
& &\times \sum_{\substack{j_1,\ldots,j_{2N-L_T}=1 \\ \sigma( j_1, \ldots, j_{2N-L_T})}}^S \prod_{k=1}^{L_T} \lambda_{j_k}^2 \prod_{l=L_T+1}^{2N-L_T} \lambda_{j_l}\,,
\en
with $L_T=L_1+L_2$. Finally, if we rearrange Eq.~\eqref{ChiChiExpansion} we find that the purity is given by the following lineal combination of elementary symmetric polynomials (Eq. (4) in the main text),
\eq
P_1 &=& \binom{N}{M}^{-1} + \frac{1}{\chi_N^2}  \sum_{L_T=0}^{N-2} \alpha_{L_T} \chi_{L_T}\chi_{2N-L_T} 
\en
where $\alpha_{L_T}$ is evaluated by recursion,
\begin{multline}
\alpha_{L_T} = \frac{(2 N-2 L_T)!}{(2 N-L_T)!}\left[\sum_{L_1=\text{Max}[0,L_T-N+M]}^{\text{Min}[L_T,M]} \frac{1}{L_1!(L_T-L_1)!} \left(\frac{M!(N-M)!}{(M-L_1)!(N-M-L_T+L_1)!}\right)^2 \right.\nonumber \\ 
 \left. - \frac{M! N! (N-M)!}{L_T! ((-L_T+N)!)^2}-\sum _{k=L_T+1}^{N-2} \alpha_k \frac{ k! (2 N-k)!}{(k-L_T)! L_T! (2 N-k-L_T)!}\right].
\end{multline}
\end{widetext}


\begin{thebibliography}{10}

\bibitem{EsteveOberthaler2008}
J.~Esteve, C.~Gross, A.~Weller, S.~Giovanazzi, and M.~K. Oberthaler,
\newblock Nature {\bf 455}, 1216 (2008).

\bibitem{Gross2010}
C.~Gross, T.~Zibold, E.~Nicklas, J.~Est{\`e}ve, and M.~K. Oberthaler,
\newblock Nature {\bf 464}, 1165 (2010).

\bibitem{RiedelTreutlein2010}
M.~F. Riedel {\em et~al.},
\newblock Nature {\bf 464}, 1170 (2010).

\bibitem{LuckeSchererEtal2011}
B.~L\"ucke {\em et~al.},
\newblock Science {\bf 334}, 773 (2011).

\bibitem{StrobelOberthaler2014}
H.~Strobel {\em et~al.},
\newblock Science {\bf 345}, 424 (2014).

\bibitem{McConnellVuletic2015}
R.~McConnell, H.~Zhang, J.~Hu, S.~Cuk, and V.~Vuletic,
\newblock Nature {\bf 519}, 439 (2015),
\newblock Letter.

\bibitem{Luo2017}
X.-Y. Luo {\em et~al.},
\newblock Science {\bf 355}, 620 (2017).

\bibitem{LuckeKlempt2014}
B.~L\"ucke {\em et~al.},
\newblock Phys. Rev. Lett. {\bf 112}, 155304 (2014).

\bibitem{PeiseKlempt2015}
J.~Peise {\em et~al.},
\newblock Nature Communications {\bf 6}, 8984 (2015).

\bibitem{TuraAcin2014}
J.~Tura {\em et~al.},
\newblock Science {\bf 344}, 1256 (2014).

\bibitem{SchmiedSangouard2016}
R.~Schmied {\em et~al.},
\newblock Science {\bf 352}, 441 (2016).

\bibitem{KilloranCramerPlenio2014}
N.~Killoran, M.~Cramer, and M.~B. Plenio,
\newblock Phys. Rev. Lett. {\bf 112}, 150501 (2014).

\bibitem{BouvrieValdesetall2016}
P.~A. Bouvrie, A.~Vald\'es-Hern\'andez, A.~P. Majtey, C.~Zander, and A.~R.
  Plastino,
\newblock Annals of Physics {\bf 383}, 401 (2017).

\bibitem{YurkeStoler1992}
B.~Yurke and D.~Stoler,
\newblock Phys. Rev. A {\bf 46}, 2229 (1992).

\bibitem{LangeKlempt2018}
K.~Lange {\em et~al.},
\newblock Science {\bf 360}, 416 (2018).

\bibitem{Fadel2018}
M.~Fadel, T.~Zibold, B.~D{\'e}camps, and P.~Treutlein,
\newblock Science {\bf 360}, 409 (2018).

\bibitem{Kunkel2018}
P.~Kunkel {\em et~al.},
\newblock Science {\bf 360}, 413 (2018).

\bibitem{Cavalcanti2018}
D.~Cavalcanti,
\newblock Science {\bf 360}, 376 (2018).

\bibitem{ChinGrimmEtal2010}
C.~Chin, R.~Grimm, P.~Julienne, and E.~Tiesinga,
\newblock Rev. Mod. Phys. {\bf 82}, 1225 (2010).

\bibitem{ChenTanLevin2005}
Q.~Chen, J.~Stajic, S.~Tan, and K.~Levin,
\newblock Physics Reports {\bf 412}, 1  (2005).

\bibitem{Giorgini2008}
S.~Giorgini, L.~P. Pitaevskii, and S.~Stringari,
\newblock Rev. Mod. Phys. {\bf 80}, 1215 (2008).

\bibitem{ZwierleinKetterle2004}
M.~W. Zwierlein {\em et~al.},
\newblock Phys. Rev. Lett. {\bf 92}, 120403 (2004).

\bibitem{BartensteinGrimm2004}
M.~Bartenstein {\em et~al.},
\newblock Phys. Rev. Lett. {\bf 92}, 120401 (2004).

\bibitem{BourdelKhaykovichEtal2004}
T.~Bourdel {\em et~al.},
\newblock Phys. Rev. Lett. {\bf 93}, 050401 (2004).

\bibitem{PartridgeEtal2005}
G.~B. Partridge, K.~E. Strecker, R.~I. Kamar, M.~W. Jack, and R.~G. Hulet,
\newblock Phys. Rev. Lett. {\bf 95}, 020404 (2005).

\bibitem{StewartJin2008}
J.~T. Stewart, J.~P. Gaebler, and D.~S. Jin,
\newblock Nature {\bf 454}, 744 EP  (2008).

\bibitem{GreinerRegalJin2003}
M.~Greiner, C.~A. Regal, and D.~S. Jin,
\newblock Nature {\bf 426}, 537 (2003).

\bibitem{CombescotMatibetEtal2008}
M.~Combescot, O.~Betbeder-Matibet, and F.~Dubin,
\newblock Phys. Rep. {\bf 463}, 215 (2008).

\bibitem{CombescotShiauChang2016}
M.~Combescot, S.-Y. Shiau, and Y.-C. Chang,
\newblock Phys. Rev. A {\bf 93}, 013624 (2016).

\bibitem{ShiauCombescotChang2016}
S.-Y. Shiau, M.~Combescot, and Y.-C. Chang,
\newblock Phys. Rev. A {\bf 94}, 052706 (2016).

\bibitem{BouvrieTichyRoditi2016}
P.~A. Bouvrie, M.~C. Tichy, and I.~Roditi,
\newblock Phys. Rev. A {\bf 95}, 023617 (2017).

\bibitem{Gaebler2010}
J.~P. Gaebler {\em et~al.},
\newblock Nature Physics {\bf 6}, 569 EP  (2010).

\bibitem{CHSH1969}
J.~F. Clauser, M.~A. Horne, A.~Shimony, and R.~A. Holt,
\newblock Phys. Rev. Lett. {\bf 23}, 880 (1969).

\bibitem{Leggett2001}
A.~J. Leggett,
\newblock Rev. Mod. Phys. {\bf 73}, 307 (2001).

\bibitem{CombescotTanguy2001}
M.~Combescot and C.~Tanguy,
\newblock EPL (Europhys. Lett.) {\bf 55}, 390 (2001).

\bibitem{CombescotShiau2015}
M.~Combescot and S.-Y. Shiau,
\newblock {\em Excitons and Cooper Pairs: Two Composite Bosons in Many-Body
  Physics} (Oxford University Pess, 2015).

\bibitem{PetrovShlyapnikov2004}
D.~S. Petrov, C.~Salomon, and G.~V. Shlyapnikov,
\newblock Phys. Rev. Lett. {\bf 93}, 090404 (2004).

\bibitem{Giorgini2005}
G.~E. Astrakharchik, J.~Boronat, J.~Casulleras, and S.~Giorgini,
\newblock Phys. Rev. Lett. {\bf 95}, 230405 (2005).

\bibitem{Nozieres1985}
P.~Nozi{\`e}res and S.~Schmitt-Rink,
\newblock J. Low Temp. Phys. {\bf 59}, 195 (1985).

\bibitem{PongLaw2007}
Y.~H. Pong and C.~K. Law,
\newblock Phys. Rev. A {\bf 75}, 043613 (2007).

\bibitem{WangLawChu2005}
J.~Wang, C.~K. Law, and M.-C. Chu,
\newblock Phys Rev. A , {\bf 72}, 022346 (2005).

\bibitem{KocikOkopiska2014}
P.~Ko\'scik and A.~Okopiska,
\newblock Few - Body Systems {\bf 55}, 1151 (2014),

\bibitem{Law2005}
C.~K. Law,
\newblock Phys. Rev. A {\bf 71}, 034306 (2005).

\bibitem{TichyBouvrie2012a}
M.~C. Tichy, P.~A. Bouvrie, and K.~M\o{}lmer,
\newblock Phys. Rev. A {\bf 86}, 042317 (2012).

\bibitem{TichyBouvrie2014}
M.~C. Tichy, P.~A. Bouvrie, and K.~M\o{}lmer,
\newblock App. Phys. B {\bf 117}, 785 (2014).

\bibitem{DalfovoStringari1999}
F.~Dalfovo, S.~Giorgini, L.~P. Pitaevskii, and S.~Stringari,
\newblock Rev. Mod. Phys. {\bf 71}, 463 (1999).

\bibitem{BrouzosSchmelcher2013}
I.~Brouzos and P.~Schmelcher,
\newblock Phys. Rev. A {\bf 87}, 023605 (2013).

\bibitem{CuestasSerra2013}
E.~Cuestas and P.~Serra,
\newblock J. Phys. B: At. Mol. and Opt. Phys. {\bf 46},
  115001 (2013).

\bibitem{SerwaneZurnEtal2011}
F.~Serwane {\em et~al.},
\newblock Science {\bf 332}, 336 (2011).

\bibitem{ZurnJochim2013}
G.~Z\"urn {\em et~al.},
\newblock Phys. Rev. Lett. {\bf 111}, 175302 (2013).

\bibitem{Jochim2015}
S.~Murmann {\em et~al.},
\newblock Phys. Rev. Lett. {\bf 114}, 080402 (2015).

\bibitem{KohstallRiedlEtal2011}
C.~Kohstall {\em et~al.},
\newblock New J. Phys. {\bf 13}, 065027 (2011).

\bibitem{TichyBouvrie2012b}
M.~C. Tichy, P.~A. Bouvrie, and K.~M\o{}lmer,
\newblock Phys. Rev. Lett. {\bf 109}, 260403 (2012).

\bibitem{BouvrieTichyMolmer2016}
P.~A. Bouvrie, M.~C. Tichy, and K.~M\o{}lmer,
\newblock Phys. Rev. A {\bf 94}, 053624 (2016).

\bibitem{EinsteinPodolskyRosen1935}
A.~Einstein, B.~Podolsky, and N.~Rosen,
\newblock Phys. Rev. {\bf 47}, 777 (1935).

\bibitem{QiongyiReid2015}
Q.~He, L.~Rosales-Z\'arate, G.~Adesso, and M.~D. Reid,
\newblock Phys. Rev. Lett. {\bf 115}, 180502 (2015).

\bibitem{WisemanDoherty2007}
H.~M. Wiseman, S.~J. Jones, and A.~C. Doherty,
\newblock Phys. Rev. Lett. {\bf 98}, 140402 (2007).

\bibitem{StewartEtal2010}
J.~T. Stewart, J.~P. Gaebler, T.~E. Drake, and D.~S. Jin,
\newblock Phys. Rev. Lett. {\bf 104}, 235301 (2010).

\bibitem{KuhnleVale2010}
E.~D. Kuhnle {\em et~al.},
\newblock Phys. Rev. Lett. {\bf 105}, 070402 (2010).

\bibitem{BergschneiderPreiss2018}
A.~Bergschneider {\em et~al.},
\newblock arXiv:1807.06405 .

\bibitem{RamanathanKurzynski2011}
R.~Ramanathan, P.~Kurzy\'nski, T.~K. Chuan, M.~F. Santos, and D.~Kaszlikowski,
\newblock Phys. Rev. A {\bf 84}, 034304 (2011).

\bibitem{CombinatoryAnalysisMacmahon}
P.~A. Macmahon,
\newblock {\em Combinatory Analysis} (Cambridge University Press, Cambridge,
  England, 1915).

\end{thebibliography}

\end{document}